\documentstyle[11pt,aaspp4]{article}

\def\etal{{\it et~al.}}
\def\eg{{\it e.g.}}
\begin{document}
\slugcomment{To appear in Experimental Astronomy}

\title{Implications of the Sample Rate on Large Space Telescopes}
\author{D.~J.~Fixsen\altaffilmark{1,2} \& R.~H.~Cornett\altaffilmark{1}}
\altaffiltext{1}{SSAI, Code 685, NASA/GSFC,
Greenbelt MD 20771}
\altaffiltext{2}{NASA Goddard Space Flight Center, Code 685, Greenbelt MD 20771}

\begin{abstract}

The frequency at which a large space telescope's (\eg NGST's) detector chips are 
read, or the sample rate, is tightly coupled to many hardware and operational 
aspects of the telescope's instrument and data handling elements. In 
this paper we discuss many of the drivers and important implications of 
the sample rate: the data rate to the ground; onboard computer storage, 
bandwidth, and speed; the number of A$\rightarrow$D chips, and therefore 
the overall size and power requirements of the analog electronics; cryocabling 
requirements; and detector noise and power. We discuss and parametrize these 
and other elements related to sample rate. Finally, we discuss the 
implications of sample rate in the context of achieving the most important 
science goals under the constraint of limited cost.

\end{abstract}

\section{Introduction}

The frequency at which each of a large space telescope's detector chips is 
read, or the sample rate, is tightly coupled to many aspects of the hardware 
aboard the spacecraft. The sample rate is an operational parameter: it is 
selected as an element of observing strategy and can be easily changed. 
However, optimum hardware design depends in many ways on the sample rate. Choosing the sample rate is therefore a critical process, and needs to be 
done early to permit hardware design to be stabilized and avoid schedule slips.
In this paper we analyze many of the hardware and operational implications and 
drivers of the sample rate. We assume the detector can be read out 
non-destructively. Destructive readouts are more properly addressed in
integration time studies.
 
The sample rate directly drives the data rate to the ground. 
This is obvious if there is no onboard processing, but still true even if 
there is data compression. The sample rate also drives several onboard
computer capability requirements, including CPU speed, bus bandwidth, and 
data storage size. The sample rate directly affects the number and performance
requirements of the A$\rightarrow$D chips, which in turn is a major 
driver of the size, weight, complexity and cost of the analog electronics. 
With the transmission line format and protocol, the sample rate determines 
the number of signal lines required between the warm electronics and the cold 
detectors, which in turn impacts the temperature and dark current of the 
detectors; in the case of instruments observing in the mid-IR (as on NGST), 
this can be a major design consideration. In addition, the sample rate drives
the read noise requirements and the final noise limits of the entire telescope,
the detector power dissipation and dark current, and, finally, determines the 
brightest objects that can be observed. Clearly, the sample rate must be 
chosen carefully, since many of the above quantities will have major impacts 
on the observatory's cost and overall performance.

In this paper we present a parameterization based on sample rate, to estimate 
and optimize observational efficiency for large space telescopes.
Our initial approach is for a generalized instrument, with applications
to NGST demonstrated in later sections.
We assume a telescope with area $A$, telescope efficiency $\epsilon$, 
detector efficiency $\eta$, and an integration time $\tau$. We will assume a 
large number of pixels $N$, and a sample rate, $R$ and calculate operational 
limits based on these parameters. We assume that the point-spread-function 
will distribute the photons across a few pixels but a large fraction $f$ of 
them will fall on a single pixel which has a full well depth of $w$.

For our discussion we have adopted the nominal "first-baseline" values for NGST 
of $A=30$~m$^2$, $\epsilon=0.7$, $\eta=0.8$, $R=0.1$~Hz $N=8\times10^7$ and $\tau=1000$~s. 
For many of the 
estimations we assume a J band filter. For a J band magnitude $B_J\approx24$ 
source the flux is 1 photon/m$^2$/sec, however, most of the results are 
simular for other bands. For many calculations it is convenient to define 
an effective area $a=A\epsilon\eta f (\sim 8$~m$^2$ for NGST).

The cost of the electronics, A$\rightarrow$D system, and backplane are all
driven by the {\it maximum} read rate. Once the cost of this rate is borne,
it is almost always an advantage to use the maximum rate. Even if it means
discarding the data it simplifies the clocking and maintains a uniform loading
on the detectors to take data at a constant rate.  Therefore, in the remainder
of the paper we will assume that the detectors are continuously read out at 
the maximum rate that the facility will support.

We have organized our discussion as follows.  In sections 2 and 3 we discuss program 
and design elements that, to first order ``prefer'' faster (\S2: Primary Drivers for 
Faster Readout) and slower (\S3: Primary Drivers for Slower Readout) sample rates. Because 
several of these elements have other, more complicated effects, in \S4 (Secondary Drivers
and Complications) we discuss both second-order effects and less dominant design elements 
affecting the sample rate, and their implications. In \S5 we discuss optimum sample rates 
for various conditions. Our conclusions are in \S5.4.

\section{Primary Drivers for a Faster Readout}
Several design considerations drive the sample rate higher. Faster sampling
leads to more reads (for the same integration) which can lower the final
noise. Faster sampling allows observing brighter objects (which could
also be accomplished with a fast shutter). Faster sampling also allows
better cosmic ray rejection, but this is only if the data are processed
to eliminate only affected samples. And for some systems faster sampling
is a way of dealing with low frequency noise in the electronics.

\subsection{Low Frequency Noise}
In the discussions here we have assumed that the noise is white; that is, it 
does not vary as a function
of frequency.  Unfortunately, the noise characteristics of the electronics
and/or detector frequently exhibit substantially higher noise at low 
frequency. This is often dubbed $1/f$ noise even when it only approximates
$1/f$ to some power. This sort of noise can sometimes be traced to the 
temperature stability of the electronics, or the power stability. Often the 
source of the instability is totally unknown or only poorly understood.
Typically, below 1~Hz (plus or minus an order of magnitude) this $1/f$ noise
is the dominate noise source.

There are several ways of dealing with this sort of noise. The most straightforward, 
running faster, unfortunately is inconsistent with the required
integration times of deep astronomical observations. In this context, one
solution is to ground the input to the first stage amplifier and take a 
``noise'' reading between each pair of normal reads. The result can be subtracted from the 
following (or preceeding) normal read. This corrects for offset drifts in
the entire chain of electronics including the A$\rightarrow$D, but at the 
cost of doubling the read rate and increasing the random read noise by a
factor of $\sqrt{2}$.

The same idea can be applied by adding a ``zero"
read every $n$ normal reads and smoothing the ``zero" reads. A natural
system is to add extra nondetector pixels at the beginning or end of each
row of the detector. An advantage of this system is that it requires no
extra circuitry and the ``zero" pixels are read out exactly as the normal
pixels are. One way to do this is to cover a one or more of the edge pixels
with an opaque layer (\eg\ aluminum) so no light gets in but in other respects
the pixels acts as a ``normal" pixel.  Since the row rate is of order 100~Hz
usually one can average several rows to obtain
a baseline to subtract from the other reads to reduce the additional noise to 
an acceptable level.

Over much of the signal band $\sim 1$~Hz to $\sim 100$~kHz systems 
have been built with relatively flat noise spectra. NGST's sheer number of pixels
($\sim10^8$) means that nearly the full bandwidth of the wires between the 
cold detector and the warm electronics will need to be utilized. Further
the long integration times ($\sim10^4$)~s will require effective treatment
of the low frequency noise. In the rest of the paper we will assume that 
the remaining noise is white. However, a full discussion of the low frequency 
noise is beyond the scope of this paper.

\subsection{Lower Noise}
At high read rates, the read noise can be averaged over many reads. If the RMS 
noise in each detector read is $\sigma$ (in photons or electrons) then the 
smallest signal that can be measured in an integration is: 
\begin{equation}
B_{min}=B_J+2.5\log({a\tau\over \sigma})+1.25\log({R\tau})
\end{equation} 
(for $2R\tau<27\sigma^2$)
if optimum Fowler sampling is used. Up-the-ramp processing improves this 
result by .06 magnitudes but the scaling remains the same.
Unfortunately the output noise of the detectors is often uncertain until
after the detectors have been tested and their performance optimized.
The output noise for a single read for NGST is uncertain as the detectors
are not yet built. Numbers of 10 to 30 e$^-$ (Fanson \etal\ 1998)
are possible. We will use $\sim30$ e$^-$ because detectors with this noise
level have been demonstrated for WFC3
(E.S. Cheng, private comunication). However, we will retain this number as a parameter
$\sigma$ so it is easy to adjust the results if lower noise
detectors are produced.

\subsection{Bright Objects}
The brighter the object, the faster the detector wells are filled, so that 
the flux of the brightest object that can be observed is directly proportional
to the rate at which the detector is read out. As the magnitude scale
is logarithmic, the bright magnitude limit is 
\begin{equation}
B_{max}=B_J-2.5\log(Rw/a)
\end{equation}
(although, of course, filters with a narrower passband or lower efficiency will
allow brighter objects). Objects this bright will fill a well in a 
single read time, so the noise will be dominated by the photon statistics or 
about $1/\sqrt{w}$, (0.4\% of the signal for $w=70000$). Other uncertainties 
(gain uncertainty, calibration stability, filter unknowns, etc) will also 
contribute, but, for bright objects a few integrations at this level will 
measure the object to the telescope's ultimate limits of precision. Presumably 
there will be other 
objects of interest in the field of view, so the observation need not stop 
after a few reads, but the remaining reads will be wasted for this bright 
object.

With minor modification the detector can observe much brighter objects (\S4.4),
but whatever rate and architecture is chosen, the brightest object will still 
be limited by the read rate.  A fast shutter would allow very bright
objects but has its own power, weight and budget costs. Since the magnitude 
scale is logarithmic, the brightest object for NGST will be within a few 
magnitudes of 20.

\subsection{Cosmic Rays}
If the data is deglitched either on-board with up-the-ramp processing 
(Fixsen \etal\ 2000) or on the ground, the cosmic ray data loss
can be limited to one or two samples after the cosmic ray event (Offenberg 
\etal\ 2000). Therefore, higher sample rates result in lower information loss.

With continuous integrations, for low-signal cases the loss of a single integration
is much more important than the loss of a single readout. The loss of continuous 
integration on average loses about 1/2 of the information, which is large 
compared to the loss of 2 to 10\% of the data for the single readout loss.

\section{Primary Drivers for a Slower Readout}
Other design considerations drive the sample rate lower. Most of these involve
cost mitigation, in the cryocables, analog electronics, onboard computer
(memory and CPU cycles), and transmitter.

\subsection{Detector Noise and Power}
The noise and power of a detector are closely coupled. High power line drivers
generate extra electrons which eventually are seen in the wells as dark current.
Read-process-generated noise is seen in the NICMOS detectors and is one of the 
limits on the NICMOS readout rate. One way to mitigate this is to use a separate
driver, but this requires an extra chip and power at the cold focus,
where space and power are at a premium. In future detectors the read noise 
will be lower and NGST read rates will be lower, but the dark current 
limits will be lower as well. This is an important relationship that warrants 
understanding as early as possible; it may drive a limit on the sampling rate, 
or it may not be a relevant driver at all.

\subsection{Cryocables}
The bandwidth of the cables from the detector to the analog electronics
is limited. Although the details are important (\S4.6) in general the 
number of wires required is:
\begin{equation}
N_W\sim20NR/\nu
\end{equation} 
For a ``standard" voltage source follower design $\nu=100$~kHz, but this depends 
on the style of analog electronics and the length of the cable. The number of 
wires may also be forced higher because of noise considerations.

The number of wires is important for several reasons. The thermal loading on 
the detectors increases as the number of wires increases. The cost increases, 
because cryocables are expensive wires in many ways. Thermal and electrical 
considerations make installation slow and laborious, since thermal and 
mechanical tie points are required on any cable. Furthermore 
the connectors and points to hold them where they can radiate without
interfering with other components become increasingly complicated with more wires. 
Finally, the low thermal conductance which is necessary requires the use of 
thin manganin wires, which are fragile and difficult to solder. This makes 
wire failure a significant probability, increasing with the number of wires.

\subsection{Analog Electronics}
To a large extent, the size, weight, power and cost of the analog electronics
is a direct function of the number of separate readout ports, which
is then tied to the speed at which a single A$\rightarrow$D can reliably
digitize $D$. 
\begin{equation}
N_{AD}=NR/D
\end{equation}
For a  practical state-of-the-art 16 bit A$\rightarrow$D, $D\approx 300$~kHz.
But, the number of A$\rightarrow$D's required is also a function of the 
A$\rightarrow$D and redundancy considerations.
Each A$\rightarrow$D requires attendant op amps, filters, sample-and-hold, 
digital drivers, and power supply equipment.

\subsection{On Board Memory}
After the data is digitized, it must be stored. We assume, as a baseline, 
that the data will be immediately processed in order to to compress it 
without loss by a factor of 2. Lossy 
compression can reduce it by another factor of 2. Fowler processing can 
compress it an additional factor of 8 to 32. Up-the-Ramp processing can 
compress it a total of $\sim64$, but requires $\sim 10$ GB of temporary 
storage (RAM). The cost of memory has been in rapid decline for many years
but GB of RAM on board a spacecraft is still expensive in terms of the
physical size, weight, and power in addition to the dollar cost of the 
RAM chips.

The baseline short term storage requirement is:
\begin{equation}
M_S=4NR\tau ~~~{\rm Byte}
\end{equation} 
which is 2 bytes per sample, and two buffers.
The long term (daily) storage requirement is:
\begin{equation}
M_L={.17NR\over K} ~~~{\rm MByte}
\end{equation}
where $K$ is the compression ratio (\S 4.9). This volume of data is 
also the volume of data that must be down-linked to the ground.
This has its own costs both on the spacecraft and on the ground. 
Of course, compression, $K$, can vary between 1 and $\sim250$.

\subsection{Computer Cycles}
The data must be handled by the onboard computer. The number of 
operations, $P$, per sample varies from $\sim6$ (to just shuffle the data 
in and out),
to $\sim45$ (to do full up-the-ramp processing with cosmic ray rejection).
With substantial compression the post-compression processing has 
a small impact because the quantity of data is much smaller.
The required processing speed is:
\begin{equation}
f_{CPU}=PNR~~~{\rm Flops}
\end{equation} 
A single 250 MFlop machine can handle the load for NGST and a modest $R$. 
However, the onboard machine may have other duties, such as, controlling the 
observatory or adjusting the optics, and more than one processor 
may be desired for redundancy.

\section{Secondary Drivers and Complications}
There are many secondary issues: calibration data formats, processing styles, 
cable length, down-link stations etc.; that also affect these numbers. Here we 
provide some explanations of how these parameters affect the above calculations.

\subsection{Calibration}
The calibration of an instrument is often as difficult as the rest of the 
observation and data reduction. For a large space telescope the problem is 
compounded by the fact that the objects convenient to view are dimmer than
standard calibration objects.

One solution is to use a dark neutral filter. In principle this can reduce a 
bright well known calibration object to a comfortable viewing brightness
even for a large space telescope. Advantages include: the time scale, absolute
brightness, and readout process are identical with the normal data processes.
This allows ``apples to apples" comparisons. There are disadvantages as
well: dark neutral filters with the required density are not easy to obtain
or verify.  Also light scattering around the dark filter can be a problem.
Finally the advantage of a short calibration observation is lost.

A second method is to use very short exposures. This has the advantage of 
changing only the time (a well understood parameter) leaving the other filters
and mirrors exactly as in the normal observations.  The disadvantage is that
the dynamic range required to compare a bright calibration object with a
``typical" field object is hard to build into the telescope. Also the 
short exposures may have different detector artifacts than the longer normal
exposures.

Finally one can use a different readout scheme (\eg\ readout a 10$\times$10
block rather than the full chip).  This enables a very high dynamic range
and also allows detailed observations of bright objects.
But the possibilities of artifacts due to the readout scheme are magnified.

Often a combination of the schemes is used, along with the development
of a series of secondary calibration sources.

\subsection{Chip Count}
After the total number of pixels is selected, there remains a question of how
many detector chips should be used. In the past the limitation was on the size
of chip that could be manufactured. But with large arrays of chips (for
NGST) there is a tradeoff between 64~1K$\times$1K chips and 
16~2K$\times$2K chips. Larger chips allow easier mounting, 
potentially fewer outputs, fewer gaps in the image area, and easier 
temperature measurement and control. 
On the other hand more chips allow more defects in a wafer, and lead to smaller
losses in the event of a failure.

\subsection{Integration Time}
The integration time is a hidden variable in many calculations. 
For high level signals, the noise is dominated by the noise inherent in the 
photons themselves. In this case, the value (statistical weight) of an 
integration is proportional to the integration time. So the integration
time is not important; the sum of several short integrations 
has the same value as if the time were spent on a single integration.

For low level signals, the situation is different. The noise is dominated
by the readout noise in the detector or other parts of the system. 
Long integration times allow the signal to peek above the noise. Thus the 
value of an integration is proportional to the square of the integration time. 
Furthermore, longer integration times allow more reads and so effectively decrease the 
noise. This is true no matter what processing is chosen. The additional number of reads
makes the full value of the integration proportional to the cube of the 
integration time. Hence, one 1000--second integration has the same statistical 
value as eight 500--second integrations when the signal is so small that
the read noise dominates the uncertainty of the measurement.

Four effects limit the integration time.

\noindent
1)Telescope drifts and instabilities make it desirable to keep the integration
times short compared to events in the observatory. For Hubble, the heating
and cooling cycle of the 80-minute orbit in some ways limits integrations to
$\sim 20$ minutes. For ground observatories similar
effects limit observations to a few hours. For NGST, similar limitations
will arise in a few months. It is important to make sure that proposed
integration times are not based on hidden assumptions of orbits or day/night
cycles.

\noindent
2)As an integration proceeds, the signal from the dark current and the
zodiacal foreground accumulates until it emerges from the readout noise.
From then on further integration adds only linearly to
the weight, so there is no advantage to a longer integration time.

\noindent
3)Eventually the wells fill up, and the detector can no longer collect 
electrons. This can happen in a part of the detector which is looking at
a bright source (\eg\ a galactic core) while another part of the detector
is collecting only a few photons (\eg\ in the fringes of the same galaxy), so the
integration time is a balance between the optimum times for the bright regions
and the faint regions.

\noindent
4)Cosmic rays strike the detector and destroy information there. The amount of
damage depends on the type of processing done to the data. Some processes
(Fowler sampling) lose all of the information in a pixel when a cosmic ray strikes.
In this case one must find a balance between losing the information already
collected and improving that information. A good analogy for this situation
is ``a bird in the hand is worth two in the bush". Other processes (including
up-the-ramp) save the information collected before the cosmic ray strike.
For these processes, the integration time can be extended to the point where
most of the pixels have been affected by cosmic rays. After that, the effective
integration time is set by the interval between incoming cosmic rays. In this
case one can chase the birds in the bush while retaining a firm grasp on the
bird already in hand.

\subsection{Readout Options}
The more complex the readout options, the more expensive (in terms of size,
power and money) the analog electronics become. The simplest approach is
to have a single fixed read rate that continuously reads through the detectors.
This has the {\em huge} advantage of allowing all of the detector clocks to
be synchronized, limiting their interference with each other. A second 
significant advantage is that A$\rightarrow$D chips and related hardware are
used much more efficiently.  However, a good
science readout rate is not sufficient to handle guiding (10 to 100 Hz is 
required).

One can handle guiding by incorporating a minor change into 
the detector chip. If the readout pixel is allowed to count in either rows 
or columns and proceed in either direction, the {\it identical} read rate can 
be used for all detectors at all times. On the guide chip, instead of 
reading out the entire chip, the readout is done several times over a region
around the guide star. For instance with a read rate of 1 rpm (.0167 Hz) on a 
2K$\times$2K chip (70 kHz) a 10$\times$10 region can be read out at 700 Hz.
Even if each read cycle starts at one corner, (0,0), the region still can
be read out at 33 Hz in the worst position (guide star on the minor 
diagonal) and at 50 Hz mean rate for random positions.

Furthermore, as long as this mode is required, one can make
a virtue of necessity and allow the mode to be used on any bright star.
A bright star can be read out immediately after resetting the detector. In 
principle this allows objects 4$\times10^6$ times (or 16 magnitudes) brighter
to be observed, which would permit NGST to observe of 4th magnitude objects.
Practically one would want a 10$\times$10 region around the star to get the 
full PSF, but this would still allow up to 9th magnitude stars to be observed
by NGST. 
(Note Jupiter is spread out over $\sim8$ M pixels, so it effectively acts as a 
constellation of 13th magnitude stars).

\subsection{Cosmic Rays}
Cosmic ray environments depend primarily on the spacecraft's orbit although
shielding can help. For example,
the cosmic ray environment at L2 is different from that in low earth
orbit; the shielding from the Earth's magnetic field and atmosphere are gone. 
But so are the concentrating effects of the Earth's magnetic field. The best 
estimates are that NSGT will experience about 1 cosmic ray per pixel every 3 
hours (Barth \& Issacs 1999). So for even modest integration times (\eg\ 1000 
sec), a significant fraction of the pixels will be affected (10\%).

\subsection{Cryocable Concerns}
One of the major undetermined parameters is how many pixels to read out per 
output. There might be several outputs per chip or only one output per chip.
The chip might be 1K$\times$1K or 2K$\times$2K (or some other size). But,
the key parameter is the readout area per output. If 1~Mpix is readout per 
output line, 160 wires are required to readout the 80 Mpix for NGST. In 
addition, each readout area needs a couple of bias settings, a clock, up/down 
control, row/column control, reset, a thermometer, and a chip reset, 
resulting in about 20 wires needed for each output.
Concern for failures due to broken wires may double this number. Thus, 80
outputs could result in up to 3200 wires. Using 2K$\times$2K blocks and not
doubling them reduces this to 400 wires. Using common grounds (for the
DC biases) and sharing the clocks and resets (within a camera) can reduce 
this to about 200 wires. A reduced wire count results in a lower thermal
load, which allows the use of larger more robust wires less likely to break.

With 640 wires per camera ($N_{cam}=4\times10^6$, with a 1K$\times$1K blocks), 
multiple connectors are required, which increases the danger of misconnected
cables and the time needed to verify that the cables are in fact working. 
However, with a 2K$\times$2K block, a single 40 pin connector (or a pair of
25 pin MDM connectors) per camera is an option.

There are advantages to fewer wires, but there are limitations. The readout 
speed is limited in a transmission line. So the number of lines required is 
directly proportional to the read rate. It is convenient to have a single 
readout per detector chip, but this not essential. A single
chip can have multiple output ports. Conversely, with an inhibit line, 
several chips could share the same output transmission line. On the warm side, 
multiplexing A$\rightarrow$D's to a single line or vice versa is 
straightforward. But debugging, noise and cross talk issues are simplified by 
an arrangement of one chip to one line to one A$\rightarrow$D.

In many tests with the readout electronics close to the detector, it has been 
convenient, given the inherent limitations of the detector output, to use
the detector output stage as a voltage follower and the readout electronics
to look at the voltage of the output. Such a system is penalized by the $RC$
time constant of the cable. In this case, the $R$ is dominated by the output
impedance of the detector, and the $C$ is dominated by the capacitance of the 
cable. About 10 time constants are needed to limit cross talk between the 
pixels to a single bit for a 16 bit system. However, the speed can be doubled
by allowing some cross talk but repairing it with post digitization processing.
If the output is used as a current source, the speed is limited by
the $L/R$ time constant. There are other noise sources in this configuration 
(current noise rather than voltage noise), but the speed is increased by about 
a factor of 5.

\subsection{Memory Options}
From the A$\rightarrow$D, the data needs to directly enter either a processor
or a memory buffer. A constant read rate allows each A$\rightarrow$D to enter 
its data into a dual port memory buffer while a processor uses the other port 
to collect and compress data from a previous read or integration. By 
synchronizing the A$\rightarrow$D with a processor (perhaps a dedicated signal 
processor) the data might be Fowler processed, and/or cross talk eliminated, 
and the requirement of a dual ported memory relaxed. If a full integration is 
required to reside in the memory, about 1 GB is required for each 2K$\times$2K 
block.

After processing, the data will need to be stored again before down-link.
It is much more important for the long term storage to be radiation-tolerant
than to be fast, since the storage will be holding data for about a day rather
than a fraction of an hour. Also errors in the raw data may be corrected
in processing, while data awaiting down-link will presumably have been compressed
and the data thus have a higher significance per bit. The down-link data rate 
depends as much on the type of processing as on the sample rate.

\subsection{Processing Options}
There are many processing options, which need not be exclusive of one another.
With a general purpose computer on board, the processing can be dynamically
adjusted to fit the data and the nature of the observations.

Fowler averaging, up-the-ramp processing, or kalman filtering can be used
to compress the data and improve the signal-to-noise ratio. One or some
combination of these may be used on most of the data.

Special purpose processing will be needed for finding the centroid of
an image for guiding. Other special processing will be required for 
spectrometers on NGST. A complex set of calculations may be 
needed on board to make the mirror adjustments. 
Telescopes that take advantage of L2 or other distant locations to limit
temperature changes will be limited in downlink data rates, which will 
make on board processing to reduce the data volume and/or make autonomous
decisions and attractive option.

\subsection{Compression Factor}
There are several types of processing that could serve to compress and/or 
deglitch the data in order to reduce the memory/transmission cost and/or 
enhance the overall observatory performance.

\noindent
1)Lossless compression is relatively cheap (in both space and CPU cycles), 
innocuous, and can lead to about a factor of 2 data compression. In some 
simulated data sets higher compression ratios have been achieved, but cosmic 
ray glitches expand the dynamic range of images and consequently reduce the 
compression ratio. Since it is cheap, lossless compression is likely to be 
used in conjunction with whatever other methods are used.

\noindent
2)Fowler processing can have large compression ratios (8 to 32), as some
or all of the reads from an integration can be combined into a single picture. 
It is simple and can be applied to the data as it arrives, reducing the need for 
short-term memory. However it does not allow cosmic ray rejection within the 
data that are processed.

\noindent
3)Up-the-ramp processing allows even larger compression ratios (16 to 64) 
since the longer integration time allowed by rejecting cosmic rays can still 
be compressed to the same single picture format. The process takes more CPU
power and more short term memory but yields higher quality data.

\noindent
4)Kalman filtering is a process that like the up-the-ramp process,
rejects cosmic rays, but one point at a time. In principle it shares the 
same advantages but requires far less memory (though more CPU cycles). 
However it is not as efficient at rejecting cosmic rays because it must use only past 
data rather than the full data set to reject the cosmic rays.

\noindent
5)Lossy compression eliminates less significant bits. If properly done little real 
information is lost, and can result in a factor 
of 2 data compression (which may be combined with the factors from the other
processes). Lossy compression is fast, but astronomers are a suspicious lot, 
loath to part with even low-significance data.

\subsection{Misconceptions}

There are some component limitations which are commonly
held to be important but which we believe to be artificial, {\em e.g.},

\noindent
The data cryocables may have always been arranged in a voltage follower
configuration, but there may be significant advantages to other arrangements.
Also the ``best" solution may use digital processing to correct for known
analog defects.

\noindent
The standard way to deal with cosmic rays is to integrate until some small
faction (5 or 10\%) of the pixels are contaminated and use several integrations
to ferret out the contaminated data. However, a detailed calculation shows that
longer integrations offer a potential increase in signal to noise, especially
when combined with a cosmic ray rejection algorithm.

\noindent
The detector designers are already doing their utmost to improve
the dark current, well depth, and output noise. But this does not preclude
minor and well understood adjustments in the chip fabrication process that would
meaningful improvements, such as allowing forward or backward clocking,
and/or row or column clocking.

\section{Optimum Speed for NGST}

\subsection{NGST Goals}

The NGST, like the HST, will be called upon to do many tasks, but it is not a 
cheap general-purpose observatory. Like Hubble, its forte will be
observing faint sources at the edge of the observable universe. While bright
stars will be observed for calibration, or for searches of their neighborhood 
for planets, observations of bright stars will be done much more cheaply
from the many observatories that exist now or that will be built 
over the next 10 years. So the $key$ consideration in the NGST's physical
and operational design is, how a process, or a component, or a decision
(about the data rate, for example) affects
the observation of faint sources at NGST's limits.

\subsection{Noise Sources and Optimizing the Sample Rate}

A key point is the integration time at which observation limits change from 
being readout noise limited to being
sky background, dark current, or cosmic ray limited. 
The background noise is the Poisson noise of the incoming photons from
the background light, which is largely reflected sunlight (for $\lambda<3~\mu$m)
or thermally emitted light from zodiacal dust (for $\lambda>3~\mu$m).
\begin{equation}
N^2_z=Z\tau
\end{equation}
Similarly dark current has its own Poisson noise:
\begin{equation}
N^2_d=I_d\tau
\end{equation}
With $\sigma$ e$^-$ noise
per readout, the readout noise for a set of uniformly sample data is:
\begin{equation}
N^2_r=12\sigma^2/R\tau
\end{equation}

\subsection{Conclusion: What is the Optimum Sample Rate?}

As the dark current and background noise $increase$ with time, and the 
readout noise $decreases$ with time, one needs only to wait long enough
and the background or dark current will be the limiting noise. However 
after some mean time $\tau$ many pixels will be affected by a cosmic ray 
and the integration can no longer continue. These formulas can be combined
to specify a lower limit for the read rate.
\begin{equation}
R>{12\sigma^2 \over (I_d+Z)\tau^2}
\end{equation}

At the present time $\sigma\sim 30$, and $\tau$ is thought to be a several thousand 
seconds for NGST's orbital environment and detectors. From COBE, $Z$ is 
about .25 photon/second for the case of NGST-like pixels, and if the dark 
current is small relative to this, the result is $R>.003$ Hz, or 
about one readout every five minutes.   This value represents a 
{\em lower~limit}
for the readout rate, as driven by known physical effects.  To allow margins 
for unknowns, and to acquire data to determine noise, it would be prudent to 
run somewhat faster than this. If lower read noise can be achieved, a slower
read rate is appropriate.

As outlined above, faster sampling has some advantages-- the potential for 
observing brighter objects, lower noise under certain conditions, and more 
complete cosmic ray rejection.  However, these advantages must be weighed 
against the important disadvantage of significantly more complicated 
electronic hardware: more and/or faster 
A$\rightarrow$D converters, with their ancillary analog electronics; 
on-board memory; more cabling (and therefore thermal complication, a serious 
issue for near- and mid-IR observatories); and more CPU power.  Meeting these
requirements will significantly multiply the overall cost of potential 
observatories.

Space telescope projects are and will continue to be strongly limited in
cost. Given this constraint it is important to treat parameter optimizations,
such as finding the optimum sample rate, as fixed in total cost. The capability
of a high sample rate, permitting observations of bright sources, may
require too many resources.  Optimization
therefore should balance science requirements against each other 
{\em at fixed cost}, with the overall goal of maximizing the science 
the observatory is designed for.

{\bf Acknowledgements}

We thank M. Greenhouse for helpful comments.

This work was supported by Lockheed-Martin.

\end{document}